\documentclass[12pt]{article}
\usepackage{amsmath}
\usepackage{amssymb}
\usepackage{amsthm}
\usepackage{psfrag}
\usepackage{graphicx}

\newcommand{\be}{\begin{equation}}
\newcommand{\ee}{\end{equation}}
\newcommand{\beqa}{\begin{eqnarray}}
\newcommand{\eeqa}{\end{eqnarray}}
\newcommand{\LL}{{\cal L}}
\newcommand{\KK}{{\cal K}}
\newcommand{\pd}{\partial}

\def\e{{\rm e}}
\def\d{\partial}
\newcommand{\bseq}{\begin{subequations}}
\newcommand{\eseq}{\end{subequations}}

\newcommand{\di}{\mathrm d}

\title{
\sc{\huge On the Extra Mode and Inconsistency of
Ho\v rava Gravity}\date{}}
\author{D. Blas,\!$^a$ O. Pujol\`as,\!$^b$ S. Sibiryakov,\!$^{a,c}$\vspace{.2cm}\\
\normalsize\llap{$^a$}
 \it FSB/ITP/LPPC,
 \'Ecole Polytechnique F\'ed\'erale de Lausanne,\\
 \normalsize\it CH-1015, Lausanne, Switzerland\\
 \normalsize\llap{$^b$}\it CERN, Theory Division, CH-1211 Geneva 23, Switzerland\\
\normalsize\llap{$^c$} \it Institute for Nuclear Research of the Russian Academy of Sciences, \\
      \normalsize \it  60th October Anniversary Prospect, 7a, 117312 Moscow, Russia}
\begin{document}

\maketitle

\begin{abstract}

  We address the consistency of Ho\v rava's proposal for a theory of
  quantum gravity from the low-energy perspective.  We uncover the
  additional scalar degree of freedom arising from the explicit
  breaking of the general covariance and study its properties.  The
  analysis is performed both in the original formulation of the theory
  and in the St\"uckelberg picture.  A peculiarity of the new mode is
  that it satisfies an equation of motion that is of first order in
  time derivatives. At linear level the mode is manifest only around
  spatially inhomogeneous and time-dependent backgrounds.  We find two
  serious problems associated with this mode. First, the mode develops
  very fast exponential instabilities at short distances. Second, it
  becomes strongly coupled at an extremely low cutoff scale.  We also
  discuss the ``projectable'' version of Ho\v rava's proposal and
  argue that this version can be understood as a certain limit of the
  ghost condensate model.  The theory is still problematic since the
  additional field generically forms caustics and, again, has a very low
  strong coupling scale.  We clarify some subtleties that arise in the
  application of the St\"uckelberg formalism to Ho\v rava's model due
  to its non-relativistic nature.

\end{abstract}

\vspace{-21.5cm} 
\begin{flushright}
\large CERN-PH-TH/2009-087
\end{flushright}

\newpage

\section{Introduction and summary}

Recently, Ho\v rava has proposed a new approach to the theory
of quantum gravity \cite{Horava:2009uw}. The key
idea of the proposal is to equip space-time with a new structure:
a foliation by space-like surfaces. This foliation defines the splitting of
the coordinates into ``space'' and ``time'' and breaks the general covariance
of general relativity (GR). Then one can improve the UV
behavior of the graviton propagator and ultimately make the
theory power-counting renormalizable by adding to the GR
action terms with higher \emph{spatial} derivatives. At
the same time the action in the ADM formalism 
contains only first order time derivatives,
which allows to circumvent the problems with the ghosts  
appearing in covariant higher order gravity theories 
\cite{Stelle:1977ry}. The higher
derivative terms naively become irrelevant in the infrared and it was
argued in \cite{Horava:2009uw} that the theory reduces to GR at large
distances.

However, the consistency of the above proposal is far from being clear.
The main concern comes from the fact that the introduction of
a preferred foliation explicitly breaks the gauge group of GR down to
the group of space-time diffeomorphisms preserving this foliation. As
already pointed out in \cite{Horava:2009uw} this breaking is expected
to introduce extra degrees of freedom compared to GR. The new degrees of
freedom can persist down to the infrared and 
lead to various pathologies (instabilities, strong
coupling) that may invalidate the theory.
An illustration of this phenomenon
is provided by theories of massive gravity where special care is
needed to make the additional degrees of freedom well-behaved 
\cite{ArkaniHamed:2002sp,Dubovsky:2004sg,Blas:2009my}.

In the recent works in the topic there have been several controversial claims about
the properties of the extra degrees of freedom.
In \cite{Charmousis:2009tc} the new mode was identified among the
perturbations around a static spatially homogeneous background in the
presence of matter. 
The mode was argued to be strongly
coupled to matter in the limit when the theory is expected to approach GR, 
making it hard to believe that a GR limit exists. It is worth noting 
that the mode found in 
\cite{Charmousis:2009tc} is not propagating: its equation of motion
does not contain time derivatives \cite{Kim:2009zn}. 
Thus it
remains unclear from this analysis whether this mode corresponds to a
real degree of
freedom or can be integrated out as unphysical.
The observation that the extra mode is non-propagating was generalized
in 
\cite{Gao:2009ht} to the case of cosmological backgrounds. The
interpretation of this result given in \cite{Gao:2009ht} is that
actually the Ho\v rava gravity is free from additional degrees of
freedom. 
It was also claimed that the strong
coupling is alleviated by the expansion of the
Universe. 
Finally, the \emph{non-linear} Hamiltonian analysis performed in
\cite{Li:2009bg} shows that the phase space of Ho\v rava gravity is
5-dimensional. This result is puzzling: a normal degree of freedom
corresponds to a 2-dimensional phase space; so the result of
\cite{Li:2009bg} suggests that the number of degrees of freedom in
Ho\v rava gravity is two and a half. Two of these degrees of freedom are
naturally identified with the two helicities of graviton. But the
physical meaning of the extra ``half-mode'' is obscure.

The aim of the present paper is to clarify this issue. We show that
Ho\v rava gravity does possess an additional light 
scalar mode. For a general
background the equation of motion of this mode contains time
derivatives implying that the mode is propagating. The peculiarity of
Ho\v rava gravity is that the equation for the extra mode is {\em first
order} in time derivatives. Still, the solution corresponds to 
  waves with a background dependent dispersion relation
 and is fixed once a single function of spatial
coordinates is determined as the initial condition
in the Cauchy problem. 
This explains why this mode corresponds to a
single direction in the phase space. 

\begin{figure}[htb]
\label{fig:1}
\psfrag{E}[][]{$E$}
\psfrag{M}[][]{$M_P$}
\psfrag{L}[][]{$\Lambda_{sc}$}
\psfrag{0}[][]{$0$}
\centerline{\includegraphics[width=0.5\textwidth]{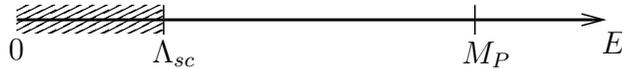}}
\caption{The energy scales appearing in the Ho\v rava model: the
  Planck mass $M_P$ and the strong coupling scale $\Lambda_{sc}$. The
  theory can be trusted only in the narrow window of energies
  $0<E<\Lambda_{sc}$ (dashed region). The scale $\Lambda_{sc}$ is much
smaller than $M_P$, it depends on the background curvature and goes to
zero for flat / cosmological backgrounds, see Eq.~(\ref{omegacut})
below.}
\end{figure}

Next we address the consistency of the Ho\v rava proposal by
  studying the infrared properties of the extra mode. We find that
  its dynamics
exhibits a number of bad
features. First, the mode becomes singular for static or spatially
homogeneous backgrounds. Namely, the mode frequency diverges in that
limit. This explains why this mode has been overlooked in the previous
analyses of perturbations in Ho\v rava gravity \cite{Charmousis:2009tc,
Kim:2009zn,Gao:2009ht}. Second, for certain
(background-dependent) values of spatial momentum the mode becomes
unstable. Again, the rate of the instability diverges if one takes the
static / spatially homogeneous limit for the background metric. 
Third, we show that at energies above a certain scale 
the extra mode is strongly coupled to
itself, and not only to matter. We find that the strong coupling scale
is background dependent and goes to zero for flat / cosmological
backgrounds. 
Hence, the model suffers from a much more severe strong coupling
problem than pointed out in \cite{Charmousis:2009tc}, where 
the dependence of the strong coupling scale on the background
curvature was ignored. Because of the strong coupling the
  Ho\v rava model can be trusted only in a narrow window of very
  small energies, way below the Planck scale; this point is
  illustrated schematically in Fig.~\ref{fig:1}.
This implies that the Ho\v rava model 
cannot be considered as consistent theory of quantum gravity.

To unveil the properties of the extra mode we make use of the
St\"uckelberg formalism. For the case at hand the St\"uckelberg trick
is synonymous to the covariantization of the model. As a result we
obtain a scalar--tensor theory with the time derivative
 of the scalar field
developing non-zero vacuum expectation value. The invariance under
foliation preserving diffeomorphisms implies that the theory 
has an internal symmetry consisting in 
reparameterizations of the scalar. We clarify
the subtleties that arise in the application of the St\"uckelberg
procedure to Ho\v rava gravity due to the intrinsically non-relativistic
nature of the proposal. The covariantization of the higher space
derivatives of the model leads to higher covariant derivative
operators in the equations of motion. Naively, this would imply the
appearance of too many degrees of freedom. However, in the theory at
hand the higher derivative operators are of a special type that allows
for a well-posed Cauchy problem with reduced number of initial data in
the preferred foliation. In this way the number of degrees of freedom
is decreased and matches with the number of modes in the 
non-covariant formulation.
As a byproduct, we point out a large class of
covariant higher derivative operators that allow for the reduction of
the number of degrees of freedom in a preferred Cauchy slicing. 

There exist two versions of the original
Ho\v rava proposal \cite{Horava:2009uw}. The difference between them
lies in an additional
restriction which can be imposed 
on the lapse function. Namely, one can require the lapse to be ``projectable'',
i.e. be constant along the foliation surfaces. In the present paper we
are mainly interested in the non-projectable case; all the
previous discussion refers to this case. The projectable 
version of the theory
is briefly discussed at
the end of the paper. We argue that in this case the Ho\v rava gravity is
equivalent to a specific limit of the ghost condensate 
model \cite{ArkaniHamed:2003uy}. 
This implies that now the theory possesses
a full-fledged extra scalar degree of freedom with second order
equation of motion. 
The classical dynamics of linear perturbations of the scalar is
regular for all backgrounds. In this sense the
projectable version of the theory is better behaved than the
non-projectable one. However, beyond the linear or classical level
the additional scalar still exhibits pathologies. As shown in   
\cite{Mukohyama:2009mz}, at the classical
level the dynamics of the projectable version of Ho\v rava gravity is
equivalent to GR supplemented by a pressureless fluid. As we discuss,
the fluid component is precisely described by the extra scalar, 
the fluid velocity being proportional to the
scalar gradient. A well-known property of pressureless fluid is
to develop caustics where the fluid velocity becomes
ill-defined. For the scalar at hand it means that the theory
inevitably breaks down after finite amount of time evolution. At the
quantum level, the extra mode exhibits unacceptably 
low scale of strong coupling. We comment on the possible ways to
address these problems. 

The paper is organized as follows. In Sec.~\ref{sec:2} 
we describe the model 
and formulate the Cauchy problem for it. In Sec.~\ref{sec:3} we derive
the linearized equations for perturbations about an arbitrary
background. We find an explicit expression for the
extra mode and show that it
obeys first order equation in time. In Sec.~\ref{sec:4} we turn to
the St\"uckelberg analysis of the model which allows us to 
study the properties of the extra mode in a transparent way.
We discuss the subtleties in the application of the St\"uckelberg
formalism to the case at hand. 
In Sec.~{\ref{sec:5}} we take the limit when the GR part decouples
from the St\"uckelberg sector and concentrate on the latter. 
We find that the linear
perturbations of the field exhibit fast exponential instability and
unacceptably low scale of strong coupling.
Finally, 
in Sec.~\ref{sec:6} we consider the projectable
version of Ho\v rava gravity 
and argue that it is equivalent to a specific
limit of ghost condensate model. 
Sec.~\ref{sec:7} contains concluding remarks and discussion of future
directions. Technical details are deferred to Appendix~\ref{app:A}.

\section{Cauchy problem for Ho\v rava gravity}
\label{sec:2}

We consider the class of non-relativistic generalizations of GR
proposed in \cite{Horava:2009uw}. One starts with the ADM
decomposition of the space-time metric,
\[
\di s^2=(N^2-N_i N^i) \di t^2-2N_i  \di x^i \di t-\gamma_{ij}\di x^i \di x^j\;.
\]
Then the action for a theory of this class
can be written in the generic form
\be
\label{ADMact}
S=\frac{M_P^2}{2}\int \di ^4x \sqrt{\gamma}\,N\,\big(K_{ij}K^{ij}-
\lambda K^2 +\xi R+\zeta R_{ij}R^{ij}+\ldots\big)\;,
\ee
where $M_P$ is the Planck mass; $K_{ij}$ is the extrinsic curvature tensor
for the surfaces of constant time, $K$ is its trace; $R_{ij}$, $R$,
$\gamma$ are the Ricci tensor, Ricci scalar and the determinant of the
spatial metric $\gamma_{ij}$, and $N$ is the lapse function. The extrinsic
curvature is related to the time derivative of the metric in the usual
way,
\be
\label{extr}
K_{ij}=\frac{1}{2N}\left(\dot\gamma_{ij}-\nabla_i N_j-\nabla_j N_i\right)\;.
\ee
Throughout the paper, if not stated otherwise, 3-dimensional indices
$i,j,\ldots$ are raised and lowered using
$\gamma_{ij}$, and the covariant derivatives carrying these indices 
are understood in the
3-dimensional sense. The ellipsis in (\ref{ADMact}) represents higher order
terms  
constructed out of the metric $\gamma_{ij}$ using only spatial
derivatives and invariant under 3-dimensional diffeomorphisms. As
discussed in \cite{Horava:2009uw}, the introduction of terms of
sufficiently high order (with six spatial derivatives) yields a theory
which is na\"ively power-counting renormalizable. 

The purpose of the present paper is to study the properties of the
theory (\ref{ADMact}) in the infrared. 
The precise structure of the higher order
terms is not essential for this analysis. 
To illustrate the general effect of these
terms it suffices to consider explicitly one of them which we choose
to be the square of the Ricci tensor.

The action (\ref{ADMact}) reduces to GR for the values of the
parameters\footnote{The value of the parameter $\xi$ is not important:
it can always be absorbed into the rescaling of the time coordinate and
the shift $N_i$.} 
$\lambda=1$, $\zeta=0$. Away from these values  (\ref{ADMact}) explicitly
breaks general covariance down to the subgroup consisting of spatial
diffeomorphisms and reparameterizations of time,
\be
\label{symm}
{\bf x}\mapsto \tilde{\bf x}(t,{\bf x})~,~~~~t\mapsto\tilde t(t)\;.
\ee
These transformations  
preserve the foliation of the space-time by the surfaces
$t=const$. Note that if one assumes that the action (\ref{ADMact})
defines a consistent quantum theory, one expects all the parameters in
this action to acquire radiative corrections and flow by the
renormalization group. In particular, the parameter $\lambda$ is
expected to be generally different from 1. It was argued 
\cite{Horava:2009uw}
that in the infrared limit the higher order terms
become negligible and so one may expect to recover GR if the parameter
$\lambda$ flows to 1 in that limit. We will see below that this
expectation is incorrect: the explicit breaking of general covariance
leads to the appearance of an extra degree of freedom in the infrared
which becomes strongly coupled\footnote{More precisely, 
the additional mode is weakly coupled only in a narrow
window at low energies. This window depends both on $\lambda$ and the
parameters of the background geometry; it
shrinks to zero both when $\lambda\to 1$ or when the background
curvature vanishes, see Sec.~\ref{sec:5}.} 
when $\lambda$ approaches $1$.

Varying the action with respect to $N$, $N_i$, $\gamma_{ij}$ yields 
the following equations,
\begin{align}
\label{Ham}
&-K_{ij}K^{ij}+\lambda K^2+\xi R+\zeta R_{ij}R^{ij}=0\;,\\
\label{mom}
&\nabla_i K^{ij}-\lambda\nabla^jK=0\;,\\
&-\frac{\d}{\d t}(K^{ij}-\lambda\gamma^{ij}K) -(1-2\lambda)NKK^{ij}
 - 2N K^{ik}K_k^j 
+N K_{kl}K^{kl}\gamma^{ij}
-\xi NR^{ij}\notag\\
&-\xi\gamma^{ij}\Delta N+\xi\nabla^i\nabla^j N 
-2\zeta NR^{ik}R_k^j-\zeta \nabla_k\nabla_l(NR^{kl})\gamma^{ij}
-\zeta \Delta(NR^{ij}) \notag\\
&+\zeta\big[\nabla_k\nabla^i(NR^{kj})+\nabla_k\nabla^j(NR^{ki})\big]=0\;.
\label{evol1}
\end{align}
Here $\Delta\equiv\gamma^{ij}\nabla_i\nabla_j$ and
we have fixed the gauge $N_i=0$.
This system has to be supplemented by the evolution equation
(\ref{extr}) for the metric which in the chosen gauge takes the form 
\be
\label{evol2}
\dot\gamma_{ij}=2N K_{ij}\;.
\ee
Let us analyze the Cauchy
problem for the system (\ref{Ham}) -- (\ref{evol2}). 
The set of initial data at $t=0$
consists of the values for $\gamma_{ij}$, $K^{ij}$ and
$N$ at this time. The initial data have to satisfy the constraints (\ref{Ham}) and
(\ref{mom}). Then Eqs.~(\ref{evol1}), (\ref{evol2})
describe the time evolution of 
the extrinsic curvature and the metric. However, the system
(\ref{Ham}) -- (\ref{evol2}) is incomplete: it does not allow to
determine the time evolution of the lapse $N$. 
In GR 
this ambiguity is a gauge artifact 
removed by appropriate gauge fixing, e.g. $N=1$. In our case 
the gauge freedom is absent, and $N$ is a genuine dynamical
field.
To obtain the missing equation 
one notices that, due to 
the lack of gauge invariance, the Hamiltonian constraint
(\ref{Ham}) is not automatically preserved by the time evolution. 
Imposing that the constraint holds at any time one gets a secondary 
constraint which produces an equation for $N$. 
Taking the time derivative of (\ref{Ham}) and simplifying the
result with the use of the rest of the equations we obtain
\be
\label{dotHam}
\begin{split}
\nabla_i\Big\{N^2\Big[\xi(\lambda-1)\nabla^i K
&+\zeta(K_{kj}\nabla^iR^{kj}-R^{kj}\nabla^iK_{kj}
 +K\nabla_j R^{ij}-R^{ij}\nabla_j K\\
 &-K_{kj}\nabla^kR^{ij}+R^{kj}\nabla_kK^i_j-K^i_j\nabla_kR^{kj}
 +R^{ij}\nabla^kK_{kj})\Big]\Big\}
=0\;.
\end{split}
\ee  
As expected, the l.h.s. of this equation vanishes identically in the
case of GR as a consequence of gauge invariance. For $\lambda$ and
$\zeta$ away from their GR values this equation allows to determine
the lapse at any moment of time 
provided the configuration of $\gamma_{ij}$, 
$K^{ij}$ is given; in this way it  
imposes additional constraint on the initial data. 
It is important
to notice that the constraint vanishes 
whenever the extrinsic curvature or gradients of the curvature
tensors are zero. In particular, this happens for spatially
homogeneous or static configurations. Note also that   
Eq.~(\ref{dotHam}) has the form of the conservation of a (space-like)
current. This fact acquires a natural interpretation in the covariant
picture where (\ref{dotHam}) becomes the equation of motion of the
St\"uckelberg field, and the current corresponds to the shift symmetry
of this field, cf. Eq.~(\ref{Jcons}) below. 

The system (\ref{Ham}) -- (\ref{dotHam}) constitutes the complete
set of
equations of motion for Ho\v rava gravity.
Let us
count the number of independent Cauchy data for this
system. Originally 
the set of initial data for 
$\gamma_{ij}, K^{ij}, N$ contains $6+6+1=13$
functions of spatial variables ${\bf x}$. 
The constraints (\ref{Ham}), (\ref{mom}),
(\ref{dotHam}) eliminate $1+3+1=5$ of them. Additionally, 3 functions are
removed by the residual (time-independent) gauge transformations of
spatial coordinates. Thus we are left with $13-5-3=5$ arbitrary
functions as initial data. 
$4$ of these functions are identified as initial data for the two
helicities of the graviton. The remaining freedom in the choice of one
more function implies the presence of an extra mode which is absent in
GR. Note that because the Cauchy data for this mode are limited to a
single function, the corresponding evolution equation must be first
order in time. This agrees with  
the
observation made in \cite{Li:2009bg} that the phase space of the Ho\v rava
gravity has odd dimensionality. 
Our task below is to
investigate the properties of the extra mode.

\section{The elusive mode}
\label{sec:3}

In this section we reveal the extra mode explicitly. 
For the sake of the argument we restrict to the case
when the higher order terms in the action are absent,
$\zeta=0$, and breaking of general covariance arises only from 
 $\lambda\neq 1$. 
This restriction allows to capture the essential physics of the extra
mode, without overloading the paper with lengthy
formulae. 
As it will become transparent below, the main conclusions are
  unaffected by the values of $\zeta$ and $\lambda$, so long as the
  general covariance is broken by some of these terms.
For $\zeta=0$ the secondary constraint (\ref{dotHam}) reduces to 
\be
\label{seccons}
\nabla_i\left(N^2\nabla^i K\right)=0.
\ee
Consider small perturbations of the fields about a 
background $\bar\gamma_{ij}, \bar K^{ij}, \bar N$. 
We assume that the background satisfies the equations of
motion of the Ho\v rava gravity, but is arbitrary otherwise. Thus we
write
\begin{align}
&\gamma_{ij}=\bar\gamma_{ij}+h_{ij}\;,\notag\\
&K^{ij}=\bar K^{ij}+\kappa^{ij}\;,\notag\\
&N=\bar N+n\;.\notag
\end{align}
The next step is to plug these expressions into 
Eqs.~(\ref{Ham}) -- (\ref{dotHam}) and expand them to linear order in the
perturbations $h_{ij}$, $\kappa^{ij}$, $n$. 
We obtain
\begin{multline}
\label{Hamlin}
\shoveright{-2\bar K_{ij}\kappa^{ij}-2\bar K^i_k\bar K^{jk}h_{ij}+
2\lambda\bar K\kappa+2\lambda\bar K\bar K^{ij}h_{ij}-
\xi\Delta h+\xi\nabla^i\nabla^j h_{ij}-\xi\bar R^{ij} h_{ij}=0\;,}
\end{multline}
\be
\label{momlin}
\nabla_i\kappa^{ij}-\lambda\nabla^j\kappa+\bar K^{kl}\nabla_k h_l^j
-\frac{1+2\lambda}{2}\bar K^{kl}\nabla^j h_{kl}
+\frac{1}{2}\bar K^{kj}\nabla_k h
+\lambda\nabla^k\bar K h_k^j-\lambda\nabla^j\bar K^{kl}h_{kl}
=0\;,
\ee
\begin{align}
&-\frac{\d}{\d t}\big(\kappa^{ij}-\lambda\bar\gamma^{ij}\kappa+
\lambda\bar Kh^{ij}-\lambda\bar\gamma^{ij}\bar K^{kl}h_{kl}\big)
-n\big((1-2\lambda)\bar K\bar K^{ij}-\lambda\bar K^2\bar\gamma^{ij}
+2\bar K^{ik}\bar K_k^j\big) \notag\\
&-(1-2\lambda)\bar N\bar K\kappa^{ij}-2\bar N\bar K^i_l\kappa^{lj}
-2\bar N\bar K^j_l\kappa^{li}
-\bar N\big((1-2\lambda)\bar K^{ij}-2\lambda\bar\gamma^{ij}\bar K\big)
\kappa \notag\\
&-\lambda\bar N\bar K^2h^{ij}-2\bar N\bar K^{ik}\bar K^{jl}h_{kl}
-\bar N\big((1-2\lambda)\bar K^{ij}-2\lambda\bar\gamma^{ij}\bar K\big)
\bar K^{kl}h_{kl} \notag\\
&+\xi\bigg[\nabla^i\nabla^j n-\bar\gamma^{ij}\Delta n -
(\bar R^{ij}-\bar\gamma^{ij}\bar R)n
+\frac{\bar N}{2}\Delta h^{ij}
-\frac{\bar N}{2}\nabla^k\nabla^i h_k^j
-\frac{\bar N}{2}\nabla^k\nabla^j h_k^i
+\frac{\bar N}{2}\nabla^i\nabla^j h\notag\\
&-\bar N\bar\gamma^{ij}\Delta h
+\bar N\bar\gamma^{ij}\nabla^k\nabla^lh_{kl}
-\frac{1}{2}\nabla_k\bar N\nabla^ih^{jk}
-\frac{1}{2}\nabla_k\bar N\nabla^jh^{ik}
+\frac{1}{2}\nabla_k\bar N\nabla^kh^{ij}\notag\\
&+\bar\gamma^{ij}\nabla_k\bar N\nabla_lh^{lk}
-\frac{1}{2}\bar\gamma^{ij}\nabla_k\bar N\nabla^kh
+\bar N\bar R^j_kh^{ik}+\bar N\bar R^i_kh^{jk}
-\bar N\bar R^{kl}\bar\gamma^{ij}h_{kl}-\bar N\bar Rh^{ij}\notag\\
&+\Delta\bar N\, h_{ij}+\bar\gamma^{ij}\nabla^k\nabla^l\bar N\,h_{kl}
-\nabla^k\nabla^j\bar N\,h^i_k-\nabla^k\nabla^i\bar N\,h^j_k\bigg]=0\;,
\label{evol1lin}
\end{align}
\begin{multline}
\label{evol2lin}
\shoveright{\dot h_{ij}=2\bar N\kappa_{ij}+2\bar N\bar K^k_jh_{ik}
+2\bar N\bar K^k_ih_{jk}+2\bar K_{ij}n\;,}
\end{multline}
\begin{align}
\shoveright{&2\nabla_i\bar K\nabla^in+\Delta\bar Kn
+2\nabla_i\bar N\nabla^i\kappa+\bar N\Delta\kappa
+\bar N\bar K^{ij}\Delta h_{ij}+2\bar N\nabla^k\bar
 K^{ij}\nabla_kh_{ij}
-\bar N\nabla_j\bar K\nabla_ih^{ij}}\notag \\
&+\frac{\bar N}{2}\nabla^k\bar K\nabla_kh 
+2\nabla^i\bar N\bar K^{kl}\nabla_ih_{kl}
+\bar N\Delta \bar K^{ij}h_{ij}-\bar N \nabla^i\nabla^j\bar
Kh_{ij}\notag \\
&-2\nabla^i\bar N\nabla^j\bar K h_{ij}
+2\nabla^i\bar N\nabla_i\bar K^{kl}h_{kl}=0\;.
\label{dotHamlin}
\end{align}
Here the indices are raised and lowered using the background metric
$\bar\gamma_{ij}$, and the covariant derivatives are understood with
respect to this metric.

One makes an important observation. As discussed above,
Eq.~(\ref{dotHamlin}) is supposed to determine the evolution of the
lapse. However, the terms linear in $n$ disappear from this equation
whenever the gradients of the background extrinsic curvature
vanish. In particular, this happens for static or spatially
homogeneous backgrounds.
 Then, instead of
determining the lapse, Eq.~(\ref{dotHamlin}) imposes a constraint
on an otherwise propagating field, making the extra mode 
non-dynamical\footnote{
The fact that the extra mode does not propagate at linear level in
homogeneous or static backgrounds also holds  
for the general Ho\v rava action
(\ref{ADMact}). 
The reason is that 
the combination in the square brackets in the secondary constraint 
(\ref{dotHam}) vanishes on these backgrounds. This is precisely the
combination which multiplies the perturbation of the lapse in the
linearized equation.}.

It is instructive to work out the case of Minkowski background in a certain 
detail. In this case
Eq.~(\ref{evol1lin}) takes the
form
\be
\begin{split}
&-\frac{\d}{\d t}(\kappa_{ij}-\lambda\delta_{ij}\kappa)
+\xi\bigg[\d_i\d_jn-\delta_{ij}\Delta n\\
&+\frac{1}{2}\Delta h_{ij}-\frac{1}{2}\d_k\d_ih_{jk}
-\frac{1}{2}\d_k\d_jh_{ik}+\frac{1}{2}\d_i\d_jh
-\delta_{ij}\Delta h+\delta_{ij}\d_k\d_lh_{kl}\bigg]=0\;.
\end{split}
\label{evol1lin1}
\ee
The linearized Hamiltonian constraint (\ref{Hamlin}) yields
$\Delta h=\d_i\d_j h_{ij}$; therefore
the trace of (\ref{evol1lin1}) reads
\[
(1-3\lambda)\dot\kappa=-2\xi \Delta n.
\]
If $n$ were determined by (\ref{dotHamlin}), this would
be a first order equation for $\kappa$. However, in Minkoswki
(\ref{dotHamlin}) reduces to 
\[
\Delta\kappa=0\;,
\]
which
restricts the perturbation of both the extrinsic curvature and the lapse to
vanish.  The rest of the argument proceeds in the same
way as in GR and one concludes that all scalar modes are
non-propagating\footnote{This statement is true only at linear order
  in perturbations. The non-linear corrections will bring back the
  propagating mode as it is clear from the study of perturbations
  in general backgrounds and from the St\"uckelberg 
  picture, see below.}. The same effect occurs in any
spatially homogeneous or static background. This explains why the extra
mode was overlooked in the previous analyses
\cite{Charmousis:2009tc,Kim:2009zn,Gao:2009ht} that focused on this class
of backgrounds.

According to the above discussion the extra mode reveals itself only
in backgrounds which are {\em both} time-dependent and spatially
inhomogeneous.
Finding an exact solution of Ho\v rava gravity with these
properties is a difficult task. Fortunately, for our purposes it is
not needed: it is enough to realize that such backgrounds exist. As a concrete
example, one can keep in mind a large gravitational 
wave\footnote{Moreover, once the perturbations to any metric are considered, 
the presence of inhomogeneities is universal.}.

In the generic case, the system of linearized equations is
intractable. 
What saves the day is the fact that the extra mode appears in
\emph{any} background such that the terms proportional
to $n$ in (\ref{dotHamlin}) do not cancel. It is enough 
to consider perturbations at space-time scales
much shorter than the characteristic distance of the variation of the
background. This allows to treat the background fields as almost
constant at the scales of interest and ensures the validity of
Fourier analysis at these scales. Technically this amounts to
keeping in the equations only terms with least number of derivatives
of the background. Let us make this point more quantitative. 
We assume that the
background metric changes at characteristic space-time scale $L$. Then
we have $\bar R_{ij}\sim 1/L^2$, $\bar K_{ij}\sim 1/L$. 
We are interested in perturbations at distances much shorter
than $L$. This means that we consider perturbations with frequencies
and momenta $\omega, p \gg 1/L$. Consequently, in Eqs.~(\ref{Hamlin})
-- (\ref{dotHamlin}) we can neglect terms with derivatives of the
background in comparison to the terms with derivatives of
perturbations. This does not imply throwing away all the terms
with background gradients: some of these terms may be the leading
ones. For instance, the first term in Eq.~(\ref{dotHamlin}) is the leading
contribution containing the lapse $n$ in this equation. In this way we
obtain the simplified system, 
\begin{align}
\label{Hamlin2}
&-2\bar K_{ij}\kappa^{ij}+2\lambda\bar K\kappa-\xi\Delta h
+\xi \pd^i\pd^jh_{ij}=0\;,\\
\label{momlin2}
&\pd_i\kappa^{ij}-\lambda\pd^j\kappa+\bar K^{kl}\pd_k h_l^j
-\frac{1+2\lambda}{2}\bar K^{kl}\pd^j h_{kl}+\frac{1}{2}\bar K^{kj}
\pd_k h=0\;,
\displaybreak[0]\\
&-\dot\kappa^{ij}+\lambda\delta^{ij}\dot\kappa
-\lambda\bar K\dot h^{ij}+\lambda\delta^{ij}\bar K^{kl}\dot h_{kl}
+\xi\bigg[\pd^i\pd^j n-\delta^{ij}\Delta n \notag\\
&+\frac{\bar N}{2}\Delta h^{ij}
-\frac{\bar N}{2}\pd^k\pd^i h^j_k-\frac{\bar
  N}{2}\pd^k\pd^j h^i_k
+\frac{\bar N}{2}\pd^i\pd^j h
-\bar N\delta^{ij}\Delta h
+\bar N\delta^{ij}\pd^k\pd^l h_{kl}\bigg]=0\;,
\label{evol1lin2}
\displaybreak[0]\\
\label{evol2lin2}
&\dot h_{ij}=2\bar N\kappa_{ij}+2\bar K_{ij}n\;,
\\
\label{dotHamlin2}
&2\pd^in\nabla_i\bar K+\bar N\Delta\kappa+\bar N\bar K^{ij}\Delta h_{ij}=0\;,
\end{align} 
where without loss of generality we have set\footnote{The
  background metric can be always brought to this form in the vicinity
  of any given point by the
  time-independent 3-dimensional diffeomorphism.} 
$\bar\gamma_{ij}\approx\delta_{ij}$.
Let us first consider Eq.~(\ref{evol2lin2}). It follows from this
equation that either $\kappa_{ij}$ or $\bar K_{ij}n$ is at least of order
$\omega h_{ij}$. Let us further assume\footnote{The analysis in the
  complementary regime $\omega\sim p$ reveals only the two transverse
  traceless modes of the graviton.} $\omega\gg p$: we will see
shortly that the dispersion relation of the extra mode obeys this
inequality. Then we see that in
Eq.~(\ref{dotHamlin2}) the last term is always negligible. 
A similar reasoning shows that 
one can neglect
all the terms containing $h_{ij}$
in (\ref{evol1lin2}). This yields a closed system of equations for
$\kappa^{ij}$ and $n$:
\begin{align}
\label{evol1lin3}
-\dot\kappa^{ij}+\lambda\delta^{ij}\dot\kappa
+\xi\big[\pd^i\pd^j n-\delta^{ij}\Delta n \big]=0\;,\\
\label{dotHamlin3}
2\pd^in\nabla_i\bar K+\bar N\Delta\kappa=0\;.
\end{align}  
We point out that this system is explicitly first order in time
derivatives. 
As already mentioned, at short scales we can treat the combinations of
the background fields appearing in the above equations as
constant. One performs the Fourier decomposition 
\[
h_{ij}, \kappa^{ij},n\propto \e^{-i\omega t+i{\bf p x}}
\]
and finds that the solution of the system (\ref{evol1lin3}) -- (\ref{dotHamlin3})
has  frequency 
\be
\label{omeg}
\omega=\frac{\xi\bar N p^4}{(1-3\lambda) p^j\pd_j\bar K}\;.
\ee
The extrinsic curvature for this solution is determined in terms of
the lapse,
\be
\label{kappij}
\kappa^{ij}=\frac{i p^k\d_k\bar K}{\bar N p^4}
\big(-(1-3\lambda)p^ip^j+(1-\lambda)\delta^{ij}p^2\big) n\;.
\ee
Note that 
$\omega\sim p^3 L^2$ which is indeed
much bigger than $p$. Besides, the extrinsic curvature behaves
as 
\be
\label{kappest}
\kappa^{ij}\sim \frac{n}{pL^2}\;.
\ee
From this estimate one concludes that the 
r.h.s. of (\ref{evol2lin2}) is
dominated by the second term. This yields for the perturbations of the
metric 
\be
\label{hij}
h_{ij}=\frac{2i(1-3\lambda) p^k\pd_k\bar K}{\xi p^4}\bar K_{ij}n\;.
\ee
Finally, we have to check the constraints (\ref{Hamlin2}),
(\ref{momlin2}).
From (\ref{hij}) we see that
\[
h_{ij}\sim \frac{n}{(pL)^3}\;.
\]
This estimate together with 
(\ref{kappest}) implies 
that the terms containing $h_{ij}$ in
(\ref{momlin2}) should be neglected compared to the first two
terms; on the other hand, all the term in (\ref{Hamlin2}) are of
the same order. Then it is straightforward to verify using the explicit
expressions (\ref{kappij}), (\ref{hij}) that the constraints are
satisfied.  

Let us briefly summarize our results. Eqs.~(\ref{kappij}) and (\ref{hij})
provide the explicit expression for the extra mode of Ho\v rava gravity in
the short wavelength limit. This mode is parametrized by a single scalar
function $n({\bf p})$ and has the dispersion relation (\ref{omeg}). Together
with the two polarizations of the graviton 
found in the complementary regime
$\omega\sim p$
this matches with our counting of degrees of freedom in
Sec.~\ref{sec:2}. 
The frequency (\ref{omeg}) of the mode diverges when the 
gradients of the background extrinsic curvature
vanish. Thus the mode becomes singular
for spatially homogeneous or static backgrounds. 

The expression 
(\ref{omeg}) also
diverges for the modes with spatial momenta perpendicular to the
gradient $\pd_i\bar K$. Naively, one could try to 
find the behaviour of these modes by applying the Fourier analysis to
the system (\ref{Hamlin2})--(\ref{dotHamlin2}) including 
next to leading term from
Eq.~(\ref{dotHamlin}), i.e. the term $\Delta\bar Kn$. 
However, this would give an incorrect result.
The leading-order expression (\ref{omeg}) for the frequency of
the mode depends on the background fields and hence on 
the space-time point where one performs the
Fourier decomposition. Therefore at the subleading level the first term in
Eq.~(\ref{dotHamlin}) produces contributions of the form
\[
 n\, t \, \pd_i \omega\pd^i \bar K \sim n\omega t L^{-3}\;,
\]
which at the time scales of interest $t\sim \omega^{-1}$ 
are of the same order as the term
\[
\Delta\bar K n\sim nL^{-3}\;.
\]
Thus, the consistent treatment of the subleading effects 
requires going beyond Fourier analysis. Instead, one has to
implement the WKB expansion in order to properly 
account for the inhomogeneity
of the background. We will perform this study in the
decoupling limit in Sec.~\ref{sec:5}, where we will find that the
subleading corrections generically lead to fast exponential
instability of the extra mode.

To close this section, let us emphasize that the qualitative
  properties of the extra mode mentioned above are
  generic for any Ho\v rava-type Lagrangian, even though in our
  consideration we mainly focused on the term 
$(1-\lambda)K^2 $.
 Technically, the important point is that the perturbation
  of the lapse $n$ enters in the linearization of the secondary
  constraint (8) with a coefficient that does not vanish on
  inhomogeneous and time-dependent 
backgrounds.  
Then this
  equation relates $n$ with the perturbation of the trace of the
  extrinsic 
  curvature $\kappa$ and leads to a first order evolution equation. 
In other words there will be always analogs of Eqs.~(\ref{evol1lin3}),
(\ref{dotHamlin3}) whenever the Lagrangian contains terms violating
the four-dimensional general covariance.
This is what makes the
  scalar mode propagate at linear level around inhomogeneous and
  time-dependent backgrounds for any Ho\v rava-type Lagrangian.

\section{St\"uckelberg formalism}
\label{sec:4}

To get more insight into the dynamics of Ho\v rava gravity
we use the St\"uckelberg formalism. This will allow us to
clearly separate the extra propagating mode and perform a detailed analysis of 
its properties.

The first step is to restore the full general
covariance\footnote{While this paper was in preparation
  Ref.~\cite{Germani:2009yt} appeared which also deals with the topic of the
  covariant form of the Ho\v rava gravity.} at the
expense of introducing the corresponding 
St\"uckelberg field. Namely, we encode the
foliation structure of Ho\v rava gravity in a scalar field $\phi(x)$ with
non-vanishing time-like gradient. The surfaces of the foliation are
then defined by the equations
\be
\label{surf}
\phi(x)=const\;.
\ee
The original action (\ref{ADMact}) is written in the gauge where the
field $\phi$ coincides with time, $\phi=t$. Below we will refer to
this choice of coordinates as ``unitary gauge''. 

Before obtaining the explicit expression for the 
action in an arbitrary gauge let us
anticipate some of its properties. First, due to the presence of the
new field $\phi$, we expect the action to be some kind of
tensor--scalar theory. Second, the invariance of the original formulation
(\ref{ADMact}) under time reparameterizations (second equation in
(\ref{symm})) translates into the symmetry of the covariant action
with respect to reparameterizations of the St\"uckelberg field,
\be
\label{repar0}
\phi \mapsto \tilde\phi=f(\phi)\;,
\ee
where $f$ is an arbitrary monotonous function. 
The appearance of a time-dependent 
vev for $\phi$ breaks the product of this symmetry and general
covariance down to the
diagonal subgroup.
The latter translates in the unitary gauge into the 
invariance under foliation preserving diffeomorphisms (\ref{symm}).

To proceed one notices that the quantities appearing in (\ref{ADMact})
are the standard geometrical objects (induced metric, 
extrinsic and intrinsic
curvature) characterizing the embedding of the hypersurfaces 
defined by (\ref{surf})
in space-time. The central object in the
construction of these quantities is the unit normal 
vector\footnote{Throughout the paper the Greek indices
  $\mu,\nu,\ldots$ are raised and lowered using 
the 4-dimensional metric $g_{\mu\nu}$ while the 
Latin indices $i,j,\ldots$ are raised and lowerd using the spatial
metric $\gamma_{ij}$. The same correspondence applies to the covariant
derivatives carrying these indices.}
$u_\mu$. Explicitly, 
\[
u_\mu \equiv \frac{\partial_\mu\phi}{\sqrt{X} }~,
\]
where
$$
X\equiv g^{\mu\nu}\,\partial_\mu\phi\,\partial_\nu\phi~.
$$
Note that $u_\mu$ is automatically invariant under the transformations
(\ref{repar0}). Other geometrical quantities associated to the
foliation are constructed out of $u_\mu$ and its derivatives. We have
the following expressions for the spatial projector:
\[
P_{\mu\nu}\equiv g_{\mu\nu}-u_\mu u_\nu\;,
\]
the extrinsic curvature:
\[
\KK_{\mu\nu}\equiv
P_{\rho\mu} \nabla^{\rho} u_{\nu}=
\frac{1}{\sqrt{X} } P_{\mu}^\rho P_{\nu}^{\sigma}\nabla_\rho
\nabla_\sigma \phi\;,
\]
and the intrinsic Riemann tensor:
\be
\label{gauss}
{\cal R}^\mu_{~\nu\rho\sigma} = 
P^\mu_\alpha \,P_\nu^\beta\, P_\rho^\gamma\, P_\sigma^\delta\, 
{}^{(4)}R^\alpha_{~\beta\gamma\delta}
+ \KK^\mu_\rho \KK_{\nu\sigma} - \KK^\mu_\sigma \KK_{\nu\rho}~,
\ee
where in the last equation ${}^{(4)}R^\alpha_{~\beta\gamma\delta}$ is
the 4-dimensional Riemann tensor. Now it is straightforward to obtain
the covariant form of the action (\ref{ADMact}) by 
identifying the quantities appearing in the ADM decomposition with 
the appropriate combinations of $u_\mu$, $P_{\mu\nu}$,
$\KK_{\mu\nu}$, etc. in the unitary gauge. For instance, in
this gauge one has 
\begin{align}
\label{uunit}
&u_0=\frac{1}{\sqrt X}=N~,~~~u_i=0\;,\\
&P^{00}=P^{0i}=0~,~~~P^{ij}=-\gamma^{ij}\;,\notag\\
&\KK_{ij}=K_{ij}~,~~~\text{etc.}\notag
\end{align}
In this way from (\ref{ADMact}) we obtain the following covariant action,
\be
\label{covar}
S=\frac{M_P^2}{2}\int \di^4x\sqrt{-g}\Big\{-{}^{(4)}R
+(1-\lambda)\KK^2+
\zeta P^{\mu\nu}P^{\rho\sigma}{\cal R}_{\mu\rho}
{\cal R}_{\nu\sigma}+\ldots\Big\}\;, 
\ee
where $\KK\equiv\KK_\mu^\mu$ and
\[
{\cal R}_{\mu\rho}=P^\alpha_\mu P^\beta_\rho\, {}^{(4)}R_{\alpha\beta}
-u^\alpha u^\beta\,{}^{(4)}R_{\alpha\mu\beta\rho}
+\KK_\alpha^\alpha\KK_{\mu\rho} 
-\KK_\mu^\alpha\KK_{\alpha\rho}\;. 
\]
In deriving (\ref{covar}) we set for simplicity $\xi=1$; as we have
already mentioned 
the value of this parameter is not physically relevant and we will
stick to this choice from now on. The above action describes gravity
interacting with a derivatively coupled scalar\footnote{Clearly, the
  terms in the action 
containing the scalar disappear when $\lambda=1$ and all the
  higher order terms vanish, i.e. in the pure GR case.} $\phi$, which enters
into (\ref{covar}) through the combinations $P_{\mu\nu}$, $\KK_{\mu\nu}$,
${\cal R}_{\mu\nu}$.

From the action (\ref{covar}) the 
advantage of 
the St\"uckelberg formalism is clear: it allows to transfer
 the extra mode of Ho\v rava gravity from the metric sector
to the $\phi$-sector. Indeed, due to the
general covariance of the action one can always choose the gauge where
the metric sector contains only
the two transverse traceless modes of the graviton.
At the same time the extra mode is unambiguously
identified with the fluctuation of the foliation structure.

At this point we encounter a puzzle. To be consistent with the
counting of degrees of freedom in the unitary gauge
the equation of motion for the St\"uckelberg field 
must be first order in time
derivatives. 
On the other hand, it is easy to see that the action (\ref{covar}) 
contains more than two derivatives of the field
$\phi$. For example, consider the term proportional to
$(1-\lambda)$. Written explicitly in terms of $\phi$ it reads,
\be\label{LL}
S_{\lambda}= \frac{M_P^2(1-\lambda)}{2}
\int \di^4x \sqrt{-g}\; \frac{1}{X} \left(\Box\phi
-\frac{\nabla^\mu\phi\nabla^\nu\phi}{X}
\nabla_\mu \nabla_\nu \phi\right)^2\;,  
\ee
and contains four derivatives.\footnote{The terms with gradients of the 
intrinsic curvature contribute with even higher derivatives of
$\phi$.} 
Thus, for general
choice of space-time coordinates, the equation of motion for $\phi$ is
fourth order in time derivatives. The corresponding Cauchy
problem requires four arbitrary initial data; this apparently 
contradicts the
counting of degrees of freedom performed in Sec.~\ref{sec:2}, where we
found only one additional function compared to GR. The resolution of this
puzzle lies in the fact that the higher-derivative equation following
from (\ref{LL}) is of a very special type. 
There exist a \emph{particular} choice of coordinates for the formulation of the
Cauchy problem where less initial data are required. This Cauchy
slicing is precisely the preferred foliation of the model.
In these coordinates the number of time derivatives in
the equation for $\phi$ is reduced to one, which  matches with the counting
of the degrees of freedom in the unitary gauge. \\

To illustrate the point about the reduction of degrees of freedom in a
preferred frame
 let us make a digression and consider the
following equation for a non-relativistic scalar $\varphi$,
\be
\label{Lifshitz}
\ddot \varphi +(-1)^q \Delta^q \varphi =0\;,
\ee
where $q\geq 2$ is an integer number. This equations describes one degree of
freedom and the corresponding Cauchy problem involves two initial
data. The general solution of the equation (\ref{Lifshitz}) is a
collection of waves with dispersion relation $\omega^2=p^{2q}$.

However, when one attempts to write down the same theory in a
manifestly Lorentz invariant way one seemingly encounters the problem
that the theory is higher derivative both in space and time
directions. Indeed, in a generic Lorentz frame Eq.~\eqref{Lifshitz}
acquires up to $2q$ time derivatives of $\varphi$. 
An observer in this frame would conclude that the number of degrees of
freedom is $q>1$, as to solve the equations of motion she 
would need up to $2q$
initial conditions. Does this contradict the counting of degrees of
freedom in the original frame? The answer is no, because the two
formulations of the Cauchy problem are {\em physically inequivalent}. 
Indeed, recall that specification of the solutions
which are considered as physical involve fixing the boundary
conditions at spatial infinity. In the simple case of
Eq.~(\ref{Lifshitz}) the natural choice is to impose vanishing of the
field,
\be
\label{condinf}
\varphi\to 0~,~~~~\text{at}~~|{\bf x}|\to \infty\;.
\ee 
When one thinks of the Cauchy problem in the boosted frame, one also
implicitly assumes the condition (\ref{condinf}) but now imposed at
the spatial infinity in this frame. Then the standard procedure is to
perform the Fourier expansion of the initial data and follow the
evolution of each eigenmode separately. In this approach one indeed
finds that in the boosted frame there are $2q$ eigenmodes. However, a
straightforward analysis shows that only two of these modes have real
frequencies. This means that the other modes grow exponentially either
at positive or at negative times. In particular, if an observer living in
the boosted frame is free to chose arbitrary initial conditions she 
will conclude that the system is unstable. However,
the growing modes, while being legitimate solutions in the boosted
frame, do not satisfy the condition (\ref{condinf}) in the original
coordinate system. Thus in the latter system they are discarded as
unphysical. 

The situation here is similar to the case of fields obeying
second-order equation of motion 
\be
\label{Lifshitz1}
\ddot\varphi-v^2\Delta\varphi=0
\ee
with superluminal velocity, $v>1$. The Cauchy problems for
Eq.~(\ref{Lifshitz1}) are physically inequivalent when formulated in
the original and highly boosted frames 
\cite{Dubovsky:2005xd,Adams:2006sv,Babichev:2007dw}. 
The latter corresponds
to the case when the Cauchy slices intersect the future causal cone
\be
\label{causcone}
t=|{\bf x}|/v
\ee  
of Eq.~(\ref{Lifshitz1}). On the other hand, all the Cauchy problems
formulated on slices lying outside the cone (\ref{causcone}) are
equivalent. The difference between Eqs.~(\ref{Lifshitz})  and (\ref{Lifshitz1})
is that the former does not have a well-defined
causal cone: the signal can propagate from the origin to any point at
$t>0$. Thus even arbitrarily small deviations from the original slicing
qualitatively change the properties of the system. 

For what follows it is convenient to slightly generalize 
the above discussion. Consider the Lorentz covariant equation
\be
\label{Lifshitz2}
A^{\mu\nu} \partial_\mu\partial_\nu \varphi -
\left(B^{\mu\nu} \partial_\mu\partial_\nu\right)^q \varphi =0\;, 
\ee
where the symmetric matrices $A^{\mu\nu}$, $B^{\mu\nu}$ 
transform as tensors under Lorentz boosts. These matrices may depend
on various fields present in the theory, in particular, they can 
depend on the field $\varphi$ itself. Naively, Eq.~(\ref{Lifshitz})
contains $2q$ time derivatives and thus describes $q$ degrees of
freedom. However this reasoning is not correct in general. 
The properties of the differential equation
\eqref{Lifshitz2} are characterized by the eigenvalues and
eigenvectors of $A^{\mu\nu}$ and $B^{\mu\nu}$. The general study of
the possible cases is beyond the scope of the present paper.
Here we just point out
the special case 
when $B^{\mu\nu}$ has one hypersurface-orthogonal 
timelike eigenvector with zero eigenvalue and
3 spacelike eigenvectors\footnote{Note that
because of symmetry of the matrix $B^{\mu\nu}$ its 
eigenvectors are orthogonal.} 
with non-zero
eigenvalues.
Then, in the frame defined by these eigenvectors the number of time
derivatives in the operator $B^{\mu\nu}\d_\mu\d_\nu$ is reduced. In
general, in curved backgrounds or 
when the matrix $B^{\mu\nu}$ is space-time dependent, the
resulting operator still contains one time derivative,
cf.~Eq.~(\ref{lapl}) below. Still, the main conclusion is that 
the number of degrees of freedom described by an equation 
can be reduced
compared to the naive expectation 
by proper choice of Cauchy slicing.\\

Let us return to the Ho\v rava gravity. Consider the higher derivative terms
appearing in the equation of motion for the St\"uckelberg field
$\phi$. To be concrete let us take the case of the action
(\ref{LL}). Then the term with four derivatives in the equation 
reads\footnote{
Explicitly in terms of 3+1 decomposition 
we have the operator identity
  \be
\label{lapl}
  P^{\mu\nu}\nabla_\mu\nabla_\nu=-\Delta +{\cal K} u^\lambda\nabla_\lambda
 \ee
valid for the action on scalar functions.}
\[
(P^{\mu\nu}\nabla_\mu\nabla_\nu)^2\phi~.
\]
This is precisely of the form (\ref{Lifshitz2}) with
$B^{\mu\nu}=P^{\mu\nu}$, $q=2$. Thus we expect a reduction of the number
of degrees of freedom in a certain frame.
In the rest of this section we demonstrate that
this is indeed the case for a large class of Ho\v rava-type
Lagrangians. Namely, we prove the following statement. Consider
perturbations of the field $\phi$ around the background $\bar\phi$, 
\be
\label{decomp}
\phi=\bar\phi+\chi\;.
\ee 
Then in the frame where the background is in unitary gauge,
\be
\label{uback}
\bar\phi=t\;, 
\ee
the linearized equation for $\chi$ is first order in
time derivative.  

For simplicity, we concentrate on the case when the Lagrangian for the St\"uckelberg
field depends only on first derivatives of the normal vector
$u^\mu$. Moreover, we assume that these derivatives enter into the
Lagrangian through the extrinsic curvature $\KK_{\mu\nu}$,
\be
\label{LKK}
{\cal L}={\cal L}(u^\mu, {\cal K}_{\mu\nu})\;.
\ee 
This case covers all terms in the general Ho\v rava-type Lagrangian
except those involving spatial derivatives of the 3-dimensional
curvature tensor $R_{ijkl}$. Indeed, the Gauss--Codazzi equation
(\ref{gauss}) implies that the terms polynomial in $R_{ijkl}$ 
depend on $\phi$ only through 
the projector $P^{\mu\nu}$ and the extrinsic curvature.

One observes that the equation of motion for the field $\phi$ has the
form of the current conservation,
\be
\label{Jcons}
\nabla_\mu J^\mu=0\;,
\ee 
where 
\[
J^\mu=\frac{\d \LL} {\d \nabla_\mu \phi} 
-\nabla_\nu \frac{\d \LL}{ \d \nabla_\mu \nabla_\nu \phi} 
\]
is the current related to the reparameterization symmetry
(\ref{repar0}). Let us demonstrate that the current is orthogonal to
the gradient of $\phi$,
\be
\label{currperp}
u_\mu J^\mu=0\;.
\ee
By explicit computation we find,
\be
\label{curr}
J^\mu=\frac{1}{\sqrt X}\left\{P^\mu_\sigma\frac{\d {\cal
      L}}{\d u_\sigma}
-u_\sigma \KK^{\mu}_\rho\frac{\d {\cal L}}{\d
  {\cal K}_{\sigma\rho}}
+P^\mu_\sigma u_\rho{\cal K}\frac{\d {\cal L}}{\d
  {\cal K}_{\sigma\rho}}
-P^\mu_\sigma P^\lambda_\rho\nabla_\lambda \frac{\d {\cal L}}{\d
  {\cal K}_{\sigma\rho}}
\right\}\;.
\ee
This expression explicitly satisfies
(\ref{currperp}). Now we perform the separation of the
field into the background and perturbations. The key observation is
that in the frame defined by Eq.~(\ref{uback}) the current contains
exactly one time derivative of the linear perturbation $\chi$. Indeed,
the perturbations of $u^\mu$ and the extrinsic curvature do not
contain time derivatives of $\chi$. This follows from the explicit
expressions
\begin{align}
&\delta u^\mu=\frac{1}{\sqrt{\bar X}}\bar P^{\mu\nu}\d_\nu\chi\;,\notag\\
&\delta {\cal K}_{\mu\nu}=\frac{1}{\sqrt{\bar X}}\Big[
-2\bar a_{(\nu}\bar P^\rho_{\mu)}\d_\rho\chi
-\bar u_\mu \bar \KK^\rho_\nu \d_\rho\chi
+\bar P^\lambda_\mu\nabla_\lambda(\bar P^\rho_\nu\d_\rho\chi)\Big]\;,\notag
\end{align}
where bar refers to the background values and 
\[
a^\nu\equiv u^\lambda\nabla_\lambda u^\nu
\]
is the proper acceleration of the congruency defined by $u^\mu$.
Then, by inspection of the expression (\ref{curr}) one finds that the
only contribution with time derivative comes from the variation of the
factor $1/\sqrt{ X}$,
\[
\delta\frac{1}{\sqrt{ X}}=-\frac{1}{\bar X}\bar
u^\sigma\d_\sigma\chi
=-\frac{1}{\sqrt{\bar X}}\dot \chi+\ldots\;.
\]
To complete the argument we have to show that taking divergence of the
current in the equation of motion (\ref{Jcons}) does not bring more
time derivatives. Due to the property (\ref{currperp}) one has
\[
J^\mu=P^\mu_\nu J^\nu\;,
\]
and thus (\ref{Jcons}) can be written as
\[
P^\mu_\nu\nabla_\mu J^\nu-a_\nu J^\nu=0\;.
\]
When expanded to linear order in perturbations, the first term contains
only spatial derivatives of the perturbation of the current. Thus this
term remains first derivative in time. Explicitly, the corresponding
contribution reads,
\[
-\frac{\bar u^\mu\bar J^\nu}{\sqrt{\bar X}}\nabla_\mu\nabla_\nu\chi\;.
\] 
Another contribution with first time derivatives comes from the
perturbation of $a_\nu$,
\[
\delta a^\nu=\frac{\bar u^\lambda\bar P^{\nu\rho}}{\sqrt{\bar X}}
\nabla_\lambda\nabla_\rho\chi+\ldots\;.
\]
Note that the two contributions are equal and sum up. This completes
the proof. \\

Two comments are in order. First, let us take a closer look at the
equation of motion for the St\"uckelberg field. For the sake of the
argument let us consider the action (\ref{LL}). Up to an irrelevant constant
factor the current
(\ref{curr}) reduces in this case to 
\[
J^\mu=-\frac{1}{\sqrt X}P^{\mu\nu}\nabla_\nu {\cal K}\;.
\]
In the unitary gauge it takes the 
form\footnote{Recall that the covariant derivatives with Latin
  indices refer to the 3-dimensional metric $\gamma_{ij}$.} 
\[
J^0=0~,~~~~J^i= N^2\nabla^i  K\;.
\]
One observes that the equation of motion 
(\ref{Jcons}) coincides with the secondary
 constraint (\ref{seccons}). It is straightforward to check that 
the equation for $\phi$ is identical to the secondary constraint also 
for the general Ho\v rava
 type Lagrangian. This confirms the consistency of the
 St\"uckelberg treatment.

Second, the expression (\ref{LKK}) is not
the most general Lagrangian compatible with the reparameterization
symmetry (\ref{repar0}). Even if one restricts attention to
Lagrangians with only first derivatives of $u^\mu$, one can still add
to (\ref{LKK}) dependence on $a^\nu$. In the unitary gauge the terms
with $a^\nu$ translate into terms with spatial derivatives of the
lapse,
\[
a^0=0~,~~~~a^i=\nabla^i N/N\;;
\]
such terms were not considered in
\cite{Horava:2009uw}. As the linear expansion of $a^\nu$ contains one time
derivative of the perturbation $\chi$, the argument given above does
not go through in this case, and the equation for the St\"uckelberg
field may contain more than one time derivative\footnote{
This is readily understood in the unitary gauge, where the
secondary constraint will now contain the time derivative of the lapse.}.
In particular, adding
the term $(a^\nu)^2$ to the Lagrangian makes the equation second order
in time. 

\section{Instability and strong coupling}
\label{sec:5}

We now perform a detailed study of the properties of the St\"uckelberg
field  perturbations
$\chi$.
To make the analysis clear we 
restrict the study to the
case when the St\"uckelberg action contains only the term
(\ref{LL}). Moreover, we consider the decoupling limit,
$(1-\lambda)\ll 1$. In this limit the backreaction of the
St\"uckelberg sector on the space-time geometry is negligible and one
considers $\phi$ as propagating in a fixed background metric. 
At the same time 
this is precisely the limit where the theory is expected to approach
GR. In this regime the non-linearities
of the St\"uckelberg
field become important at energy scales much smaller than those of the other modes
in the theory. This will allow to easily establish the strong coupling
scale in this sector.

We start by obtaining the quadratic
action for the perturbations. According to the discussion of the
previous section we work in the foliation defined by the background
value of the field. Thus we fix the gauge $\bar\phi=t$, $\bar
N_i=0$. Expanding the quantities in the action (\ref{LL}) to quadratic order
one obtains
\be
\label{quadract}
\begin{split}
S_\chi=\frac{M_P^2(1-\lambda)}{2}\int \di^4x\sqrt{\gamma}\;
\bigg[&-2\bar N^2\nabla_i\bar K\dot\chi\nabla^i\chi
+\bar N^3(\Delta \chi)^2\\
&-2\bar N\nabla_i\nabla_j \bar N^2\,\nabla^i\chi\nabla^j\chi
+\left(\frac{2}{3}\Delta \bar N^3
-\bar N^2\dot{\bar K}\right)\nabla_i\chi\nabla^i\chi\bigg]\;.
\end{split}
\ee
where we have used 
\begin{align}
&{\cal K}=\bar K-2\nabla_i\bar N\nabla^i\chi-\bar N\Delta \chi
+\bar N\dot\chi\Delta\chi
+2\bar N\nabla^i\dot\chi\nabla_i\chi\notag\\
&~~~~~~~+2\nabla_i\bar N\dot\chi\nabla^i\chi
+\dot{\bar N}\nabla_i\chi\nabla^i\chi
+\frac{\bar N^2}{2}\bar K\nabla_i\chi\nabla^i\chi
-\bar N^2\bar K^{ij}\nabla_i\chi\nabla_j\chi\;.\notag
\end{align}
This action explicitly reveals the properties of the extra mode
discussed previously.
Indeed, it
contains only one time derivative of $\chi$, so that the resulting
equation is first order in time. The term with time
derivative vanishes whenever the gradient of the extrinsic curvature
is zero, and the field $\chi$ becomes 
non-propagating.
One also observes that due to the background
equation of motion (\ref{seccons}) this action is invariant
under the shifts
\[
\chi\mapsto \chi+\xi(t)\;,
\]
where $\xi(t)$ is an arbitrary function of time. This is 
recognized as 
the linearized form of the reparameterization (\ref{repar0}).
This symmetry 
prevents $\chi$ from having the ordinary $\dot\chi^2$ kinetic term.

Let us analyze the equation of motion following from 
(\ref{quadract}). As in
Sec.~\ref{sec:3} we are interested in the short wavelength limit,
\[
pL\gg 1\;,
\] 
where $p$ is the momentum of the $\chi$-mode, and $L$ is the
characteristic length of the variation of the background. In this
regime the dominant terms in the equation are those
containing the largest number of derivatives of $\chi$.  After
using the equations for the background (\ref{seccons}), the leading
contributions to the equations of motion
are
\be
\label{eqq}
2\nabla^i \bar K\d_i\dot\chi+\bar N\Delta^2\chi
+6\nabla^i\bar N\nabla_i\Delta\chi=0\;,
\ee 
where for future reference we retain the main subleading
correction represented by the last term on the l.h.s. 
To find the solution of Eq.~(\ref{eqq}) 
we use the same strategy as in
Sec.~\ref{sec:3}. Namely, we restrict to the vicinity of a given point 
$x_o$. To the leading approximation the background fields in this
vicinity can be 
considered as constant. One also assumes that the spatial metric
$\bar\gamma_{ij}$ at this point is flat and the Cristoffel symbols
vanish; this can be always achieved by performing a 3-dimensional 
diffeomorphism. Now it easy to obtain the leading behavior of the
solution. 
One performs   
the Fourier expansion  
\be
\label{plain}
\chi\propto \e^{-i\omega t+i{\bf px}}
\ee
and substitutes it into (\ref{eqq}). Discarding the last -- subleading
-- term one obtains the
dispersion relation 
\be
\label{omegnew}
\omega=-\frac{\bar Np^4}{2(p^i\d_i\bar K)}\;.
\ee
This coincides with the expression\footnote{Recall that in the present
  section we work within the convention $\xi=1$.} (\ref{omeg}) of
Sec.~\ref{sec:3} in the decoupling limit $\lambda\approx 1$.

We now discuss the first corrections in $1/(pL)$ to the solution
(\ref{plain}), (\ref{omegnew}). The reason for considering these
corrections is that they 
qualitatively change the behaviour of the mode making it either
exponentially decaying or growing. 
As already pointed out in Sec.~\ref{sec:3}, in order to find the
subleading corrections one has to go beyond the Fourier analysis and
implement the WKB expansion in the vicinity of the point $x_o$. The
details of this procedure are contained in the Appendix. Here we quote
the main result. The eigenmode frequency acquires an imaginary part
which is estimated as
\be
\label{deltaomeg}
\delta\omega\sim ip^2L\;.
\ee
The sign of this imaginary part depends on the direction of the mode
momentum ${\bf p}$ relative to the gradients of the background. Those
modes for which the imaginary part is positive are exponentially
growing. Note that the rate of this growth is much faster than the
characteristic background frequency $1/L$. 
Thus we conclude that the Ho\v rava model
suffers from fast instability at short scales.

Another problem with the theory appears when one takes into account
self-interaction of the field $\chi$. Let us first consider the case
of flat background. Then, computing the leading non-linear terms from
the expansion of the covariant Lagrangian (\ref{LL}) in the
perturbation $\chi$ we obtain,
\be\label{int}
S = \frac{M_P^2(1-\lambda)}{2}\int \di^4x 
\left\{ 
{(\Delta\chi)^2}
+2 { \dot\chi \left( (\Delta\chi)^2 +2\partial_i\chi\partial_i\Delta\chi \right)}
\right\}\;.
\ee
As expected, the quadratic part of the action does not contain time
derivatives. Note, however, that the time derivatives do appear in the
interaction. 
The form of this Lagrangian 
is restricted by the fact that $\chi$
nonlinearly realizes the field-reparameterization symmetry (\ref{repar0})
\[
\chi \mapsto \chi + \xi(t) + \dot\xi(t) \chi + \dots\;.
\] 
The theory \eqref{int} clearly has a dimensionful coupling constant
$(\sqrt{|1-\lambda|}\,M_p)^{-1}$ which signals the presence of strong
quantum coupling at high enough energies. The naive estimate of the
strong coupling scale is provided by the inverse of this coupling
\cite{Charmousis:2009tc},
\be
\label{naivecut}
\Lambda_{naive} = \sqrt{|1-\lambda|} \; M_P~.
\ee
Note that this scale goes to zero in the putative GR limit $\lambda\to
1$. In particular, in this limit it is parametrically smaller
  than the ``deep UV'' scale $M_P$ at which the
  higher-order terms of the Ho\v rava model could become important. 

We now argue that the scale of strong coupling for the action
(\ref{int}) is even lower than (\ref{naivecut}), viz. zero. The
physical reason is that due to the absence of time derivatives in
the quadratic part of the action, rapid fluctuations of the field
$\chi$ are not suppressed. Hence, the interaction terms with time
derivatives blow up (see related discussion in
\cite{Nicolis:2008in}).

To make a quantitative statement let us regulate the action
(\ref{int}) by expanding on a nearly flat but nontrivial
background. This introduces the first order kinetic term for $\chi$ as
in (\ref{quadract}). 
For momenta much larger than the scale defined
by the background the leading order terms in the action are essentially the same as
in flat space plus the background dependent kinetic term.
Schematically,
\[
S = \Lambda^2_{naive} \int \di^4x 
\left\{ 
L^{-2}\; v^i\dot\chi \partial_i \chi + {(\Delta\chi)^2}
+ { \dot\chi (\Delta\chi)^2 }
\right\}\;,
\]
where $v^i$ is the unit vector along the direction of the extrinsic
curvature gradient, and
$L$ is the typical length scale of the background. 
The free part of the action is invariant under the 
scaling\footnote{Remarkably, 
we find here the same relative scaling of space and time
as that proposed by Ho\v rava for the deep UV to make the theory naively
power-counting renormalizable. 
We have nothing more to
say on that coincidence.}
\begin{align}
&x\mapsto b^{-1} \, x \cr\notag
&t\mapsto b^{-3} \, t \cr\notag
&\chi\mapsto b \, \chi ~.\notag
\end{align}
 Under this scaling the
interaction term has dimension $+4$; thus it becomes relevant at short
scales. To estimate the cutoff one performs the rescaling 
\begin{align}
&t\mapsto\hat t=tL^2\;,\notag\\
&\chi\mapsto\hat\chi=L^{-1}\Lambda_{naive} \chi\;,\notag
\end{align}
which brings the quadratic part of the action to the canonically
normalized form. This yields,
\[
S=\int \di^4x\left\{v^i\dot{\hat\chi}\d_i\hat\chi+(\Delta\hat\chi)^2
+\frac{L^3}{\Lambda_{naive}}\dot{\hat\chi}(\Delta\hat\chi)^2 
\right\}\;,
\]
where dot is now understood as the derivative with respect to the
rescaled time $\hat t$. From this expression one reads out the cutoff
scales for the spatial momentum and the rescaled frequency: they are
set by the appropriate powers of the unique coupling constant
appearing in the interaction term. One obtains
\begin{align}
&\Lambda_p=L^{-3/4}\Lambda^{1/4}_{naive}\;,\notag\\
&\Lambda_{\hat\omega}=L^{-9/4}\Lambda^{3/4}_{naive}\;.\notag
\end{align}
Using the relation $\hat\omega=\omega/L^2$ between the rescaled and
the physical frequencies we obtain the physical frequency
cutoff
\be
\label{omegacut}
\Lambda_{\omega}=L^{-1/4}\Lambda^{3/4}_{naive}\;.
\ee
Note that due to the non-relativistic structure of the theory the
cutoff scales in spatial momentum and frequency are not
equal. Instead, they satisfy the relation 
$\Lambda_\omega=\Lambda_p^3L^2$, which is compatible
  with the dispersion relation of the extra mode.
It is clear that both $\Lambda_p$ and $\Lambda_\omega$ go to zero in
the limit of flat background. Hence, the theory becomes strongly coupled
at all scales.

\section{Ho\v rava gravity with projectable lapse as a ghost
  condensate}
\label{sec:6}

In this section we consider another version of Ho\v rava theory also
proposed in \cite{Horava:2009uw}. In this formulation the lapse
function appearing in the action (\ref{ADMact}) is assumed to be
projectable, which means that it does not
depend on spatial coordinates, $N=N(t)$. As we are about to see, the
dynamics of the model in this case is very different from the case 
studied in the main body of the paper. We discuss it here only briefly,
leaving a more thorough study for the future.

In the projectable case the variation of the action with respect to $N$,
instead of producing the local Hamiltonian constraint (\ref{Ham}),
gives the integral of Eq.~(\ref{Ham}) over the whole space. Such an
integral constraint does not affect local physics. Thus, as far as 
local dynamics is concerned, the full set of equations of motion is
provided\footnote{Clearly, Eq.~(\ref{dotHam}) expressing the
  conservation of the Hamiltonian constraint is also absent.} 
by Eqs.~(\ref{mom})-(\ref{evol2}). Using the
reparameterization of time one can set $N=1$ in these equations. Let us
count the number of required Cauchy data. Out of 12 functions present
in $\gamma_{ij}$, $K^{ij}$, 3 are constrained by Eq.~(\ref{mom}). 3
more functions are removed by residual spatial diffeomorphisms. Thus
we are left with $12-3-3=6$ initial conditions. This is larger by 2
than in the GR case, suggesting the presence of an extra
mode with second order evolution equation.

The simplest way to study this new mode is by using the St\"uckelberg formalism. First, 
notice that the constraint $N=1$ can be enforced by adding to the
action (\ref{ADMact}) the term
\[
S_\rho=\int \di^4x\sqrt{\gamma}N\,\frac{\rho}{2}\left(\frac{1}{N^2}-1\right)\;, 
\]  
where $\rho$ is a Lagrange multiplier. From the relation (\ref{uunit}), 
the generally covariant form of the previous action reads
\be
\label{Srho1}
S_\rho=\int \di^4x\sqrt{-g}\,\frac{\rho}{2}
\big(\nabla_\mu\phi\nabla^\mu\phi-1\big)\;.
\ee
Apart from this term the action for $\phi$ contains higher-order contributions
coming from the rest of the Ho\v rava Lagrangian. The latter
contributions either contain
more than two derivatives or describe non-minimal couplings
to the metric, cf. Eq.~(\ref{covar}). 

The theory (\ref{Srho1}), together with higher order
contributions, can be viewed as a special (non-minimally coupled) 
case of the 
 ghost condensate
model \cite{ArkaniHamed:2003uy}. 
The latter model is characterized by a
Lagrangian 
\be
\label{ghcond}
{\cal L}_{gc}=P(\nabla_\mu\phi\nabla^\mu\phi) + (\text{terms with
  higher derivatives})\;, 
\ee  
where the function $P(X)$ has a minimum at $X=1$. This forces the
field $\phi$ to develop a non-zero time-like gradient. The Lagrangian
(\ref{Srho1}) is the ``sigma model limit'' of (\ref{ghcond}) when the
function $P$ gets replaced by the constraint. Note, however, an
important difference in the interpretation of the higher derivative
terms in the ghost condensate model and Ho\v rava gravity. In the
standard approach to the ghost condensate they are considered as
higher order terms in the effective field theory expansion, while
according to Ho\v rava they should be taken at face value and determine
the UV properties of the system.

Let us get more insight into the dynamics of the field $\phi$
described by (\ref{Srho1}). For the sake of the argument we omit the
higher order terms. This would correspond to the IR limit of the
putative UV complete theory. However, we will see shortly that the
$\phi$-sector exhibits a number of pathologies which make the UV
completion problematic. 

The action yields the following equations of motion,
\begin{align}
\label{unitgr}
&\nabla_\mu\phi \nabla^\mu\phi=1\;,\\
&\nabla_\mu(\rho\nabla^\mu\phi)=0\;.\notag
\end{align}
These equations are equivalent to the equations of motion of an ideal
pressureless fluid (dust) with energy--momentum tensor
\[
T_{\mu\nu}=\rho\,\nabla_\mu\phi\nabla_\nu\phi\;.
\]
We see that $\rho$ and $\nabla_\mu\phi$ are identified respectively as
the energy density and velocity of the effective fluid. This result is
in agreement with the findings of \cite{Mukohyama:2009mz} where it was
proposed to interpret the effective dust-like component arising in
the projectable version of the Ho\v rava gravity as dark matter. However,
we point out that a well-known property of pressureless fluids is that they 
develop caustics, i.e. there are space-time regions where the fluid
velocity 
is ill-defined (see \cite{Felder:2002sv} 
for discussion of this topic in the context of field theory models of
dark matter). 
The formation of caustics is easy to understand in the decoupling
limit when the backreaction of the fluid 
on the geometry is negligible. Then
the fluid particles move along
geodesics without feeling each other. Given a general inhomogeneous
initial distribution of particle velocities, their trajectories will
cross forming a caustic. While this process does not pose a
problem for real dust where it leads to virialization, the formation of
caustics means an inconsistency in the case of a scalar theory, as the field
 is not differentiable at the caustics. Thus, we conclude that for generic
initial configurations the theory described by the action
(\ref{Srho1}) breaks down after finite time evolution. Note that the
formation of caustics is a general problem of the ghost condensate action
(\ref{ghcond}) \cite{ArkaniHamed:2005gu}. In that context, it was suggested 
\cite{ArkaniHamed:2005gu}
that the problem might be resolved by the effect of higher derivative
term which make the fluid particles deviate from the geodesic
motion. This hope is absent in the special case of the action
(\ref{Srho1}): irrespective of the higher-order terms 
the fluid particles move exactly 
along geodesics as long as the
constraint (\ref{unitgr}) is present. 

The consistency of the Lagrangian
(\ref{Srho1}) is also challenged  at the quantum
level. To get the flavor of the problem
let us again omit the higher order terms and
proceed  to quantize the model in the  canonical formalism. For
simplicity we
also neglect the backreaction of the field $\phi$ on the metric
  and assume the latter to be flat.
The canonically
conjugate momenta for the variables $\phi$, $\rho$ are
\begin{align}
&\pi_\phi=\rho\dot\phi\;,\notag\\
\label{pirho}
&\pi_\rho=0\;.
\end{align}
The constraint (\ref{unitgr}) in canonical variables takes the form
\be
\label{ugr1}
\frac{\pi_\phi^2}{\rho^2}-(\d_i\phi)^2=1\;.
\ee
Equations (\ref{pirho}) and (\ref{unitgr}) form a pair of second class
constraints and enable to eliminate the  variables $\rho$,
$\pi_\rho$. In this way one obtains the Hamiltonian,
\be
\label{HH}
{\cal H}=\pi_\phi\sqrt{1+(\d_i\phi)^2}\;.
\ee
Note that in our case the elimination of constraints does not modify
the Poisson brackets of the remaining variables,
\[
\{\pi_\phi({\bf x}), \phi({\bf y})\}=\delta({\bf x}-{\bf y})~,~~\text{etc.}
\]
The quantization of the theory proceeds now in the standard way by
imposing the canonical commutation relations on $\pi_\phi$ and $\phi$.  

The Hamiltonian (\ref{HH}) certainly looks unusual. To get insight
into its properties, let us expand it  around the
background $\phi=t$, $\pi_\phi=\rho_0$. Thus we write
\[
\phi=t+\chi/\sqrt{\rho_0}~,~~~~\pi_\phi=\rho_0+\pi_\chi\sqrt{\rho_0}
\]
and obtain,\footnote{Note that the canonical transformation from the
  variables $\phi$, $\pi_\phi$ to $\chi$, $\pi_\chi$ is
  time-dependent. Taking this properly into account eliminates the
  term linear in $\pi_\chi$ in the Hamiltonian.}
\be
\label{HH1}
{\cal H}=\frac{1}{2}(\d_i\chi)^2
+\frac{1}{2\sqrt{\rho_0}}\pi_\chi (\d_i\chi)^2+\ldots\;.
\ee
As an example we included here one of the interaction terms. Now,
(\ref{HH1}) describes a theory with dimensionful coupling
$1/\sqrt{\rho_0}$. This implies that the theory gets strongly coupled
at the cutoff scale $\Lambda\lesssim \rho_0^{1/4}$. For the dark matter
interpretation of the $\phi$-sector, taking the present-day value for
the average
density of dark matter, one obtains the cutoff
\[
\Lambda \lesssim 10^{-3} \mathrm{eV}\;,
\] 
which is unacceptably low for a candidate theory of quantum
gravity. In fact, the cutoff in the theory (\ref{HH1}) 
may be even lower, viz. zero. This can be argued
from the fact that the quadratic part of the
Hamiltonian (\ref{HH1}) does not contain the momentum $\pi_\chi$. Thus
the quantum fluctuations of $\pi_\chi$ are not suppressed and the
interaction terms containing $\pi_\chi$ blow up. The careful analysis of this 
issue is beyond
the scope of this paper\footnote{Another issue which we do not address
in this paper is the effect of the higher order terms on the
power-counting.}.  

To summarize, we found that 
in the decoupling limit, when the backreaction of the extra mode
  on the metric can be neglected, the projectable version of the 
Ho\v rava gravity  
suffers from caustic and low cutoff problems. This suggests the
inconsistency of the model in its present form.

\section{Conclusions and discussion}
\label{sec:7}

In this paper we have studied Ho\v rava's proposal for quantum gravity
\cite{Horava:2009uw} from the low-energy perspective. We have mainly
concentrated on the ``non-projectable'' version of the model.
We have
uncovered the additional scalar degree of freedom arising from the
explicit breaking of the general covariance and analyzed its
properties in detail. A peculiarity of the new mode is that it
satisfies an equation of motion that is of first order in time
derivatives; this means that it adds just one direction to the phase
space, or `half' a degree of freedom. At linear level the mode is
manifest only on spatially inhomogeneous {\em and} time-dependent
backgrounds. 
We have demonstrated the existence of the extra mode in two
  physically equivalent
  ways. First we performed the analysis in the original ADM-like
  formulation of Ho\v rava and identified the mode among the
  perturbations of the metric. Second, we made use of the
  St\"uckelberg formalism which amounted to restoring the general
  covariance of the model. This required introducing a scalar field
  describing the foliation structure of the Ho\v rava model. It is
  worth stressing that this procedure did not add any new physical
  degrees of 
  freedom compared to the original formulation because the scalar
  field was introduced simultaneously with a new local symmetry. The
  latter gauge symmetry is the part of general covariance which is
  explicitly broken in the Ho\v rava's original formulation. The
  St\"uckelberg approach allowed us to transfer the extra mode into
  the fluctuations of the foliation structure and to perform a
  detailed study of its properties.

We found
two 
serious problems associated with this mode. First, the mode develops
very fast exponential instabilities at short distances. Second, it
becomes strongly coupled at extremely low cutoff scale. 
Due to the non-relativistic nature of the theory the cutoff scales in
spatial momentum and frequency are different. They both depend on the
curvature of the background metric and go to zero when this curvature
vanishes. 
These features
allow to conclude that Ho\v rava's proposal is inconsistent in the
present form.

We have also discussed the ``projectable'' version of the Ho\v rava
model. 
We have argued that this version  can be understood as 
a certain limit of ghost condensation. 
In this case, the theory propagates
a whole degree of freedom with second order equation in time for any
background. 
In this sense the projectable version is better behaved than the
non-projectable one.
However, the preliminary analysis of the model in the regime when
the backreaction of the extra mode on the space-time geometry can be
neglected suggests that the model
is still problematic since the additional field  
generically forms 
caustics and, again, has very low strong coupling scale.

Let us comment on the possible ways in which the problems of the
theory can be
addressed. The comparison of projectable and
non-projectable cases suggests that
the only way to make the extra mode well-behaved is to promote it to a
full-fledged scalar. There are two strategies to do so.  One
possibility is to relax the requirement of invariance under
reparameterizations of time and reduce the symmetry group of the
theory down to  
(time-dependent)
spatial diffeomorphisms.
As a result one obtains
some version of the ghost condensate model \cite{ArkaniHamed:2003uy}
which is known to be a consistent effective theory up to relatively high
scales. The problem with caustics
can be also addressed in this framework \cite{ArkaniHamed:2005gu}.
In
the unitary gauge the theory would be described by a generalization
of Ho\v rava action containing 
arbitrary 
functions of the lapse $N$. Needless to say, it is unclear if the
appealing UV properties of the Ho\v rava proposal can be preserved
within this approach. An argument suggesting that it may be
impossible is the break-down of the black hole thermodynamics in
the ghost condensate model \cite{Dubovsky:2006vk,Feldstein:2009qy},
and more generally, in 
theories with low-energy violation of Lorentz invariance 
\cite{Eling:2007qd}.
Indeed, the break-down
of thermodynamics indicates violation of unitarity in the underlying
theory, which makes construction of a renormalizable and hence UV
complete quantum theory having the ghost condensate as its low-energy
limit problematic.

Alternatively, one can promote the extra mode to a whole scalar 
without breaking the invariance under
foliation-preserving diffeomorphisms. 
To achieve this one has to add to Ho\v rava's Lagrangian terms that
are 
non-linear in $N$ and respect the symmetry. An example of such term is
$N^{-2}(\nabla_i N)^2$.
In the St\"uckelberg picture around flat background 
this corresponds to the addition
of the term $(\partial_i \dot\chi)^2$.  In this case also the equation for
the St\"uckelberg field 
$\chi$ becomes of second order in time for any background. Still, the
terms added to the Lagrangian in the St\"uckelberg language
are higher order in total number of derivatives. 
It remains an open question if this feature does not lead to any
pathologies. 
The problems with the break-down of black hole thermodynamics
would arise in this approach as well:
 the field $\chi$ generically has
different propagation velocity compared to the graviton (helicity-2
excitation), thus violating the Lorentz invariance at low energies. It
is unclear at present if and how these problems can be resolved.

Finally, we mention another interesting outcome of our
study. We observed that the equation of motion
for the St\"uckelberg field in
Ho\v rava gravity has a peculiar structure. This equation is
explicitly higher order in covariant derivatives. Consequently
in a general
coordinate frame it is higher order in time. Generically, this
would imply the presence of additional ghost modes. However, this
reasoning is incorrect in our case: 
there
is a unique preferred frame where the number of time derivatives in
the equation is drastically reduced. Solving the Cauchy problem in
this frame requires less initial data than in a general frame. The
difference can be traced to the intrinsically non-relativistic nature
of the theory which implies that the two frames are physically
inequivalent. Technically this manifests itself in the fact that the 
boundary conditions at spatial infinity are not equivalent in the two
frames. We pointed out that the reduction of degrees of freedom in a
preferred Cauchy slicing is
generic for a wide class
of higher order covariant derivative operators. This opens up the
possibility to construct a new class of consistent 
higher derivative theories with equations of motion based on these
operators. 
We leave the investigation of these issues for the future. 

\paragraph*{Acknowledgments}

We thank Gia Dvali, Cristiano Germani, Riccardo Rattazzi, 
Valery Rubakov and Oleg Ruchaysky for
stimulating discussions. We are grateful to Luis \'Alvarez-Gaum\'e, 
Alexey Boyarsky, Robert
Brandenberger,  Dmitry
Gorbunov, Michele Redi and Alex Vikman for encouraging interest. 
We thank Shinji Mukohyama for useful comments on the first
  version of this paper.
This work was supported in part by the Swiss Science Foundation
(D.B.), the Tomalla Foundation (S.S.), RFBR grant 08-02-00768-a (S.S.)
and the Grant of the President of Russian Federation
NS-1616.2008.2 (S.S.).

\appendix
\section{WKB expansion for the extra mode}
\label{app:A}

In this Appendix we construct the solution of Eq.~(\ref{eqq}) using
the WKB method. The application of this method is legitimate in the
case when the frequency $\omega$ and momentum $p$ of the field $\chi$
are much larger than $L^{-1}$, where $L$ is the characteristic
space-time scale of the background. The leading-order approximation
to the solution is obtained in the main text and is given by the plain
wave (\ref{plain}) with the frequency related to the momentum by the
dispersion relation (\ref{omegnew}). Our aim here is to obtain the
order $1/(pL)$ corrections to this solution.

We work locally in the vicinity of the point $x_o$, which without loss
of generality we assume to coincide with the origin of the coordinate
frame. By the 3-dimensional diffeomorphism we can make the spatial
metric at this point flat, and its first derivatives vanish. 
This implies that  the spatial covariant derivatives in
Eq.~(\ref{eqq}) can be replaced by ordinary ones in our
approximation. 
We now use the
following ansatz for the  field $\chi$:
\be
\label{ans}
\chi\propto\exp\Big[-i\omega^{(0)}t+ip_ix^i
-i\delta\omega t+i\frac{a}{2}t^2+ib_itx^i+\frac{i}{2}c_{ij}x^ix^j\Big]\;.
\ee 
Here $\omega^{(0)}$ is the leading-order frequency (\ref{omegnew})
evaluated at $x_o$; $\delta\omega$ is the frequency shift; and $a$,
$b_i$, $c_{ij}$ are the coefficients in the Taylor expansion of the
WKB phase at $x_o$. Note that if we want the ansatz (\ref{ans}) to 
represent a small correction to the plain wave solution
(\ref{plain}) in the region $t,x^i\ll L$, we must require
\be
\label{assump}
\delta\omega\ll\omega^{(0)}~,~~~~~a\sim\omega^{(0)}L^{-1}~,~~~~~
b_i, c_{ij}\sim pL^{-1}\;.
\ee
One proceeds by evaluating the derivatives of $\chi$ appearing in
Eq.~(\ref{eqq}).  Keeping the first subleading corrections one
obtains,
\begin{align}
\label{didt}
&\d_i\dot\chi=\big[p_i\omega^{(0)}+p_i\delta\omega-p_iat-p_ib_jx^j
+\omega^{(0)}b_it+\omega^{(0)}c_{ij}x^j+ib_i\big]\chi\;,\\
\label{dD}
&\Delta^2\chi=\big[p^4+4p^2p_ib_it+4p^2p_ic_{ij}x^j
-4ip_ip_jc_{ij}-2ip^2c_{ii}\big]\chi\;.
\end{align}
The last term in Eq.~(\ref{eqq}) is already subleading because of the
additional derivative of the background. Thus, it is enough to
evaluate the corresponding $\chi$-derivative to the leading order
\be
\label{diD}
\d_i\Delta\chi=-ip_ip^2\chi\;.
\ee
Finally, the inhomogeneity of the background is taken into account by expanding
the coefficients in (\ref{eqq}) in a Taylor series,
\begin{align}
\label{dKexp}
&\d_i\bar K=\d_i\bar K_{o}+\d_i\d_j\bar K_o x^j+\d_i\dot{\bar K}_ot\;,\\
\label{Nexp}
&\bar N=\bar N_o+\d_i\bar N_o x^i+\dot{\bar N}_ot\;,
\end{align}
where the quantities with the subscript ``o'' are evaluated at
the origin $x_o$. We will omit this index in what follows.

The next step is to substitute the expressions (\ref{didt}) --
(\ref{Nexp}) into Eq.~(\ref{eqq}) and require the vanishing of the Taylor
coefficients up to linear order in coordinates. This yields the
following system of equations,
\begin{align}
\label{abc1}
&-2 p_i\d_i\bar K a+\big(2\omega^{(0)}\d_i\bar K+4\bar Np^2p_i\big)b_i
+2\omega^{(0)}p_i\d_i\dot{\bar K}+p^4\dot{\bar N}=0\;,\\
\label{abc2}
&-2 p_i\d_i\bar K b_j+\big(2\omega^{(0)}\d_i\bar K+4\bar Np^2p_i\big)c_{ij}
+2\omega^{(0)}p_i\d_i\d_j\bar K+p^4\d_j\bar N=0\;,\\
\label{abc3}
&\ p_i\d_i\bar K \delta\omega
+i\d_i\bar K b_i-i\bar N(2p_ip_jc_{ij}+p^2c_{ii})
-3ip^2 p_i\d_i\bar N=0\;.
\end{align}
Note that within the assumptions (\ref{assump}) the first term in
Eq.~(\ref{abc2}) is much smaller than the rest and we can neglect
it. From this equation, the solution for $c_{ij}$ reads:
\[
c_{ij}=-\frac{V_iV_j}{U_k V^k}\;,
\] 
where 
\begin{align}
&U_i=2\omega^{(0)}\d_i\bar K+4p^2p_i\bar N\;,\notag\\
&V_j=2\omega^{(0)}p^i\d_i\d_j\bar K+p^4\d_j\bar N\;.\notag
\end{align}
It is easy to check that $c_{ij}$ satisfies the estimate
(\ref{assump}). We do not need to solve the system
(\ref{abc1})--(\ref{abc3}) in the full generality:
 any special solution is sufficient
for our purposes. Thus we set $b_i=0$ and find $a$  from 
(\ref{abc1}).
 Finally, from Eq.~(\ref{abc3}) we obtain the frequency
shift
\[
\delta\omega=\frac{i}{p^i\d_i\bar K}\left\{
-\frac{2\bar N(p_jV^j)^ 2}{U_k V^k}-\frac{\bar Np^2V_jV^j}{U_k V^k}+3p^2p_j\d_j\bar
N\right\}\;. 
\]
This expression is purely imaginary. Analyzing the order of magnitude
of the terms entering into it one obtains the estimate
(\ref{deltaomeg}) used in the main text.


\end{document}